\documentclass[12pt]{article}
\usepackage{epsfig}

\begin{document}
{\noindent \it \small Eur. Phys. J. B}
\begin{center}
{\Large \bf Sandpiles on fractal bases: \\
pile shape and phase segregation}

\vskip 1.0cm
{\large \bf N.Vandewalle$^{(a)}$ \footnote{corresponding author, e-mail:
nvandewalle@ulg.ac.be}, R.D'hulst$^{(b)}$}
\vskip 1.0cm

$^{(a)}$ {\it GRASP}, Institut de Physique B5, Universit\'e de Li\`ege, \\
B-4000 Li\`ege, Belgium.
\vskip 0.5cm
$^{(b)}$ Dept. of Mathematics and Statistics, Brunel University,
Uxbridge, Middlesex UB8 3PH, London, UK.

\end{center}


\newpage
{\noindent \large Abstract}
\vskip 0.6cm

Sandpiles have become paradigmatic systems for granular flow studies in
statistical physics. New directions of investigations are discussed here.
Rather than varying the nature of the pile (sand, salt, rice,..) we have
investigated changes in the boundary conditions. We have investigated
experimentally and numerically sandpile structures on bases having a
fractal perimeter. This type of perimeter induces the formation of a quite
complex set of ridges and valleys. A screening effect of the valleys is
observed and depends on the angle of repose. Binary granular systems have
also been investigated on such bases: a spectacular demixing is obtained
along the valleys. This type of phase segregation is discussed with respect
to numerical studies.

\newpage
{\noindent \large 1. Introduction}
\vskip 0.6cm

Despite their everyday familiarity, sandpiles and granular flows have
become paradigmatic systems of new complexity problems in physics
\cite{bakbook,herrmann}. Granular matter shows behaviors that are
intermediate between those of solids and liquids: powders pack like solids
but flow like liquids. However, granular flows are non-Newtonian
\cite{bagnold}. Thus, dry and wet granular frictions are a practical and
theoretical challenge \cite{barabasi}.

The most basic property of sandpiles is certainly the angle of repose
\cite{jaeger}, i.e. the angle $\theta$ made between the horizontal and the
apparent surface of the pile. This angle can take values between $\theta_r$
(the angle below which the pile is stationary) and $\theta_d$ (the angle
above which avalanches flow down the surface). In between $\theta_r$ and
$\theta_d$, the sandpile manifests some bistability: it can be either
stationary or in a state of avalanches.

The symmetry of the base on which the sandpile is constructed is a relevant
parameter which can sometimes lead to exotic pile shapes. When the base is
a disk (Figure 1a), the pile has a conic shape; every point of the surface
being characterized by the angle of repose $\theta$. When the base is a
square (see Figure 1b), a pyramidal pile is usually obtained. Again, every
point of the surface is characterized by a local angle $\theta$. However,
ridges emerge within a four-fold structure (see the grey lines in Figure
1b). It should be noticed that the ridges are characterized by an angle
less than $\theta$. For more complex bases having a constant convexity, it
has been proven that the pile is a so-called {\it tectohedron} geometrical
object \cite{iss} for which all facets are inclined with similar $\theta$
angles. The facets meet on a network of ridges which can describe the pile
shape. When the base has a low symmetry, various distinct states exist for
the pile shape (see the grey lines in Figure 1c). Up to now, the case of
non-convex bases has not received much attention.

Intuitively, one can imagine that for non-convex bases, the pile exhibits
numerous valleys in addition to the network of ridges. The situation is
then more complex.

In order to import other basic symmetry conditions, it seems of interest to
consider fractal-like systems. This should lead to a wide variety of length
scales as well as simple power laws for characterizing physical properties
\cite{feder}.

In section 2, we focus our experimental investigations on the effects of
such bases with a fractal perimeter as the one illustrated in Figure 1d.
This will underline thereby the interest for studying such sandpiles and
the effect of the boundary conditions on the pile structure. We also
present some experiments with binary mixtures of sand. In section 3, some
simulation work is presented allowing some qualitative interpretation of
our findings.

\vskip 1.0cm
{\noindent \large 2. Experiments}
\vskip 0.6cm

Figure 2 presents a picture of a sandpile built on the base illustrated in
Figure 1d. The perimeter of such a base results from 3 iterations of a
generator and has a fractal dimension close to $D_f = {\ln{5} \over
\ln{3}}$. The arrows point to a large valley and inner sub-valleys. The
fractal perimeter implies that a hierarchy of valleys emerges on the sides
of the pile.

One could first ask if well placed holes in the inner part of the base
instead of a fractal perimeter can produce the same phenomenon. The answer
is that the formation of valleys can be created using holes in the base but
the valleys and ridges are not especially hierarchically distributed. A
fractal perimeter allows for the study of ``natural" structures. These
structures look like natural erosion patterns which have been also
recognized as being fractal objects \cite{erosion}.

Other sandpiles on various bases with different fractal dimensions $D_f$
have been numerically and experimentally investigated. The pictures are not
shown here because in all cases, a hierarchical structure of valleys and
ridges is formed. However, depending on the repose angle $\theta_r$ of the
pile, different shapes are observed on the same fractal base. It can be
understood that the pile extension is small and avalanches are mainly
dissipated on the valleys close to the center of the base if the sand has a
high repose angle. Thus, only a limited part of the fractal perimeter
participates to the dissipation dynamics or ejection of grains when $\theta
> \theta_r$. However, with a low repose angle, the pile is large and the
whole fractal perimeter participates to the dissipation of avalanches.

Consider further the case of a binary sand mixture. It is known that the
difference in the repose angles of two kinds of particles can produce phase
segregation inside the sandpile \cite{demixing}. Demixing is often observed
but depending on various parameters like (i) the input flow, (ii) the grain
size and (iii) the morphology of each species. A spectacular
self-stratification can be sometimes produced. This has also been proven
experimentally \cite{demixing}, numerically \cite{makse} and theoretically
\cite{degennes}. However, this self-stratification observed in a vertical
Hele Shaw cell cannot be visualized from the exterior of a tri-dimensional
sandpile.

However, on bases with a fractal perimeter, the situation is quite
different. As an example, a mixture of dark and white sand grains has been
used: (i) white grains: with diameter between 0.2 -- 0.3 mm and (ii) brown
grains: with diameter between 0.07 -- 0.1 mm. The two species have
different angles of repose. The angles $\theta_r$ of brown and white grains
have been estimated to be $38^o$ and $32^o$ respectively. We have checked
that this mixture leads to an internal self-stratification as observed in a
vertical Hele Shaw cell. When this mixture is poured on a classical regular
base (a disk), no specific structure is observed on the conic surface.
However, on the base of Figure 1d, a phase segregation is clearly observed
along the valleys (see Figure 3). Indeed, valleys remains in dark and
ridges are in white. Also in the subvalleys, the phase segregation is
observed. Due to the fractality of the perimeter, the phase segregation
results in alernating vertical strates. Thus, the phase segregation can be
visualized from the pile exterior itself for sandpile having a fractal-like
perimeter. This effect is relevant in e.g. industry where granular piles do
not have especially a conic shape.

\vskip 1.0cm
{\noindent \large 3. Simulations}
\vskip 0.6cm

In order to find out some information on the geometrical structure made of
valleys and ridges, we have performed numerical simulations of pile shapes.
These simulations are based on the common cellular automaton which has been
extensively discussed in the relevant literature \cite{btw,automata}. In
this model, the sandpile is built on a two-dimensional lattice, where the
grains occupy only a single lattice site. Here, $h_{i,j}$ denotes the
height of the sandpile at coordinate $i,j$. At each time step, an arbitrary
number $N$ of grains is dropped at the top of the central column of the
lattice. Then, the pile is relaxed as follows. The dynamics of each grain
at position $i,j$ on the sandpile surface is governed by the four local
angles of repose: $\tan^{-1}{|h_{i,j}-h_{i+1,j}|}$,
$\tan^{-1}{|h_{i,j}-h_{i-1,j}|}$, $\tan^{-1}{|h_{i,j}-h_{i,j+1}|}$, and
$\tan^{-1}{|h_{i,j}-h_{i,j-1}|}$. When one or several of these angles is
larger than the repose angle $\tan^{-1}{(z_c)}$, some grains are assumed to
roll on the surface until they reach a stable configuration where the local
slope is less than the repose angle. The ``rolling grain" follows downward
or/and sideways random paths on the pile surface. Figure 4 presents a
tri-dimensional sketch of the top of the pile where a single grain is
deposited and rolls down until it reaches a stable local configuration. The
simulation stops when the sandpile reaches one site on the border of the
base and when the pile shape is then nearly invariant.

The model is thus based on the classical sandpile models encountered in the
scientific literature since the introduction of the Bak-Tang-Wiesenfeld
model \cite{btw}. The new ingredient we introduce in view of the
observation in section 2 is the shape of the base which is herein
considered to have a fractal perimeter. In addition to the perimeter
geometry, only one parameter controls the pile shape: $z_c$. For
convenience, only integer values of $z_c$ have been considered in this
work. Lattice sizes up to $200 \times 200$ have been used for the larger
bases.

A typical example of the network of ridges that is numerically obtained is
illustrated in Figure 5. The base is the one of Figure 1d. The structure is
markedly branched and reaches a high level of complexity. The main features
of the simulated structures are recognized to be those of the experimental
sandpiles: both valleys and ridges exist and are hierarchically
distributed.

One should note that the network of ridges of Figure 5 is calculated with a
low $\theta_r$ value ($z_c=2$) in order to reach all parts of the
perimeter. The partial screening effect has deep analogies with the
diffusion of entities through fractal interfaces studied by Sapoval and
coworkers \cite{sapoval}.

The amazing phase segregation described in section 2 can be also simulated.
The above numerical model can be generalized to the case of two distinct
species. Following the Head and Rodgers model \cite{head}, four parameters
should then be considered: $z_c^{\alpha \beta}$ corresponds to the maximum
slope on which a particle of type $\alpha =1,2$ can remain on the top of a
particle $\beta=1,2$ without starting to roll down. A typical set of
parameter values giving self-stratification is: $z_c^{11}=5$, $z_c^{12}=4$,
$z_c^{21}=3$ and $z_c^{22}=3$. Figure 6 presents the top view of a pile
obtained with this set of parameter values. The binary sandpile was built
on a base with a fractal perimeter. In Figure 6, a phase segregation is
clearly observed at proximity of the holes, i.e. near the valleys. This is
consistent with experimental results.

However, it is not yet clear how to choose the relevant parameters
$z_c^{\alpha \beta}$. Indeed, the diagonal elements $z_{\alpha \alpha}$ of
the $z_c$-matrix are obviously the angles of repose of each pure species
$\alpha$. These parameters can be measured in the macroscopic world.
However, the non-diagonal elements of $z_c$ do not have a macroscopic
counterpart. It should be noted that some sets of parameter values lead to
``exotic" patterns which are not encountered in the experiments.

In the light of our numerical results, the phase segregation is interpreted
as follows. The species having the largest angle $\theta_r$, i.e. the brown
species, cannot reach the whole perimeter areas, before the other type,
whence avalanches are dissipated in the primary valleys. Also, the complete
network of ridges of Figure 4 cannot form. The white species has however a
low angle of repose and reaches the borders of the base. Thus, primary
valleys are controlled by the largest angle of repose and secondary valleys
and ridges are controlled by the second angle of repose leading
approximately to the picture of dark valleys and white ridges as in Figure
3.

\vskip 1.0cm
{\noindent \large 4. Conclusion}
\vskip 0.6cm

In summary, we have investigated experimentally sandpiles on bases with a
fractal perimeter. The shapes of the sandpiles exhibit a hierarchical
(fractal) structure of valleys and ridges. Phase segregation has been found
to be vizualized. The repose angle $\theta_r$ seems to be the fundamental
parameter since $\theta_r$ allows or not the pile to take the whole fractal
shape or not. Moreover, we have shown that lattice models provide useful
numerical tools for describing qualitatively the phenomenology that we
discovered herein. More precise experimental studies should be made in the
future.

Our work also suggests new directions of investigations. Instead of varying
the nature of the pile (sand, rice,...) \cite{ricepile}, we have suggested
hereabove to change the conditions on the boundaries where avalanches are
dissipated. Theoretical investigations of these are fascinating goals.

\vskip 1.0cm
{\noindent \large Acknowledgements}
\vskip 0.6cm

NV is financially supported by the FNRS. A grant from FNRS/LOTTO allowed to
perform specific numerical work. Fruitful discussions with E.Bouchaud,
S.Galam, E.Clement and J.Rajchenbach were appreciated.

\newpage
{\noindent \large Figure Captions}

\vskip 1.0cm
{\noindent \bf Figure 1} --- Illustration of bases on which sandpiles are
built: (a) a circular base; (b) the four-fold network of ridges of the pile
build on a square base; (c) irregular convex polygon and its network of
ridges; (d) a base having a fractal perimeter ($D_f = {\ln{5} \over
\ln{3}}$). The perimeter of the bases is drawn in black while the network
of ridges is denoted in grey.

\vskip 1.0cm
{\noindent \bf Figure 2} --- Sandpile made on the fractal base as in Figure
1d. The diameter of the sand grains is in the range 0.2 -- 0.3 mm. Valleys
and subvalleys are indicated.

\vskip 1.0cm
{\noindent \bf Figure 3} --- Sandpile on the fractal base as in Figure 1c.
A mixture of two kinds of sand grains has been used: (i) diameter of white
grains: 0.2 -- 0.3 mm and (ii) diameter of brown grains: 0.07 -- 0.1 mm.
Phase segregation (demixing) is clearly observed along the valleys.

\vskip 1.0cm
{\noindent \bf Figure 4} --- Tri-dimensionnal sketch of the rule for the
present cellular automaton model. Each grain is deposited at the top of the
pile and rolls down following the relaxation rule. The parameter is herein
$z_c=3$.

\vskip 1.0cm
{\noindent \bf Figure 5} --- Top view of a numerical simulation of the
network of ridges obtained for a granular pile on the base as in Figure 1d.

\vskip 1.0cm
{\noindent \bf Figure 6} --- Top view of a typical configuration of the
binary pile using the parameters $z_c^{11}=5$, $z_c^{12}=4$, $z_c^{21}=3$
and $z_c^{22}=3$.

\newpage
\begin{picture}(21,27)(40,300)
\epsfig{file=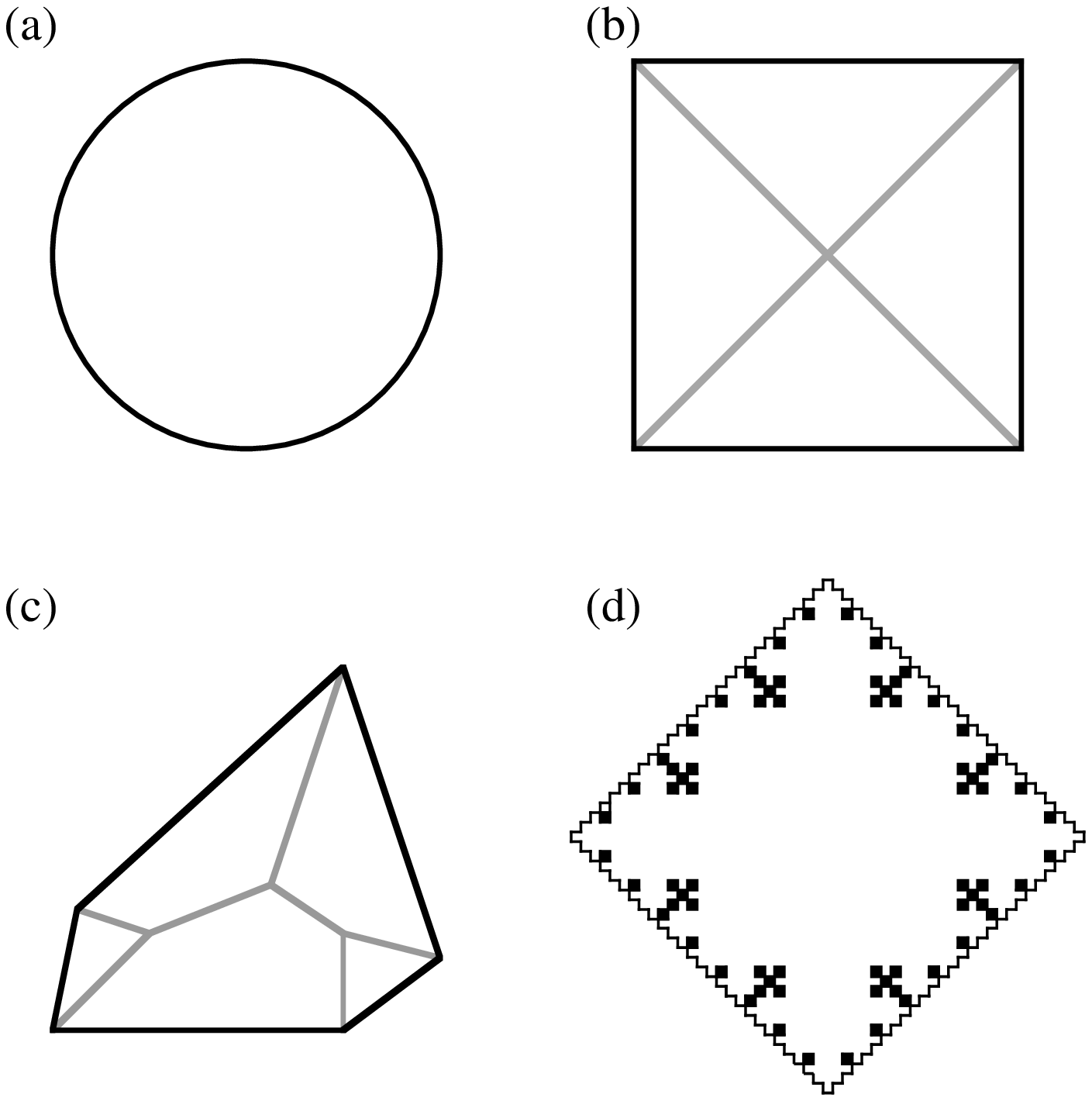}
\end{picture}
\newpage

\begin{picture}(21,27)(3,200)
\epsfig{file=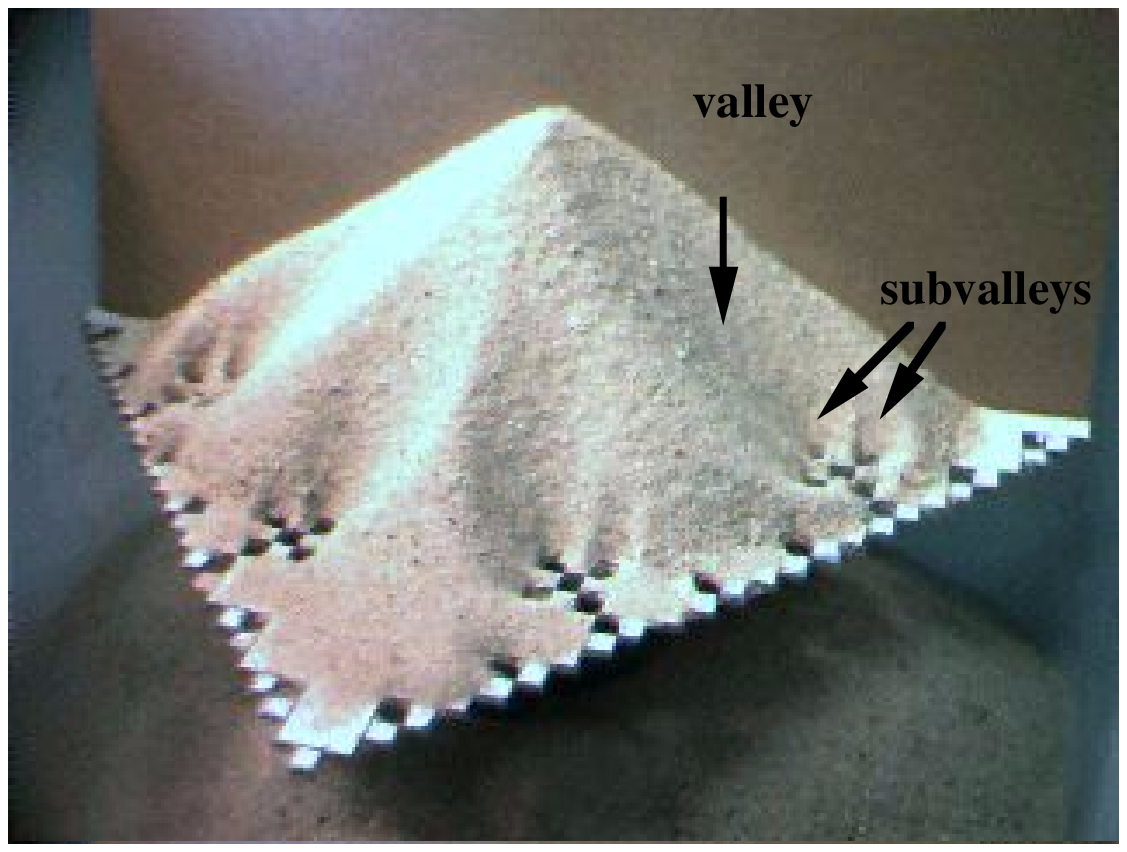,angle=90}
\end{picture}
\newpage

\begin{picture}(21,27)(3,200)
\epsfig{file=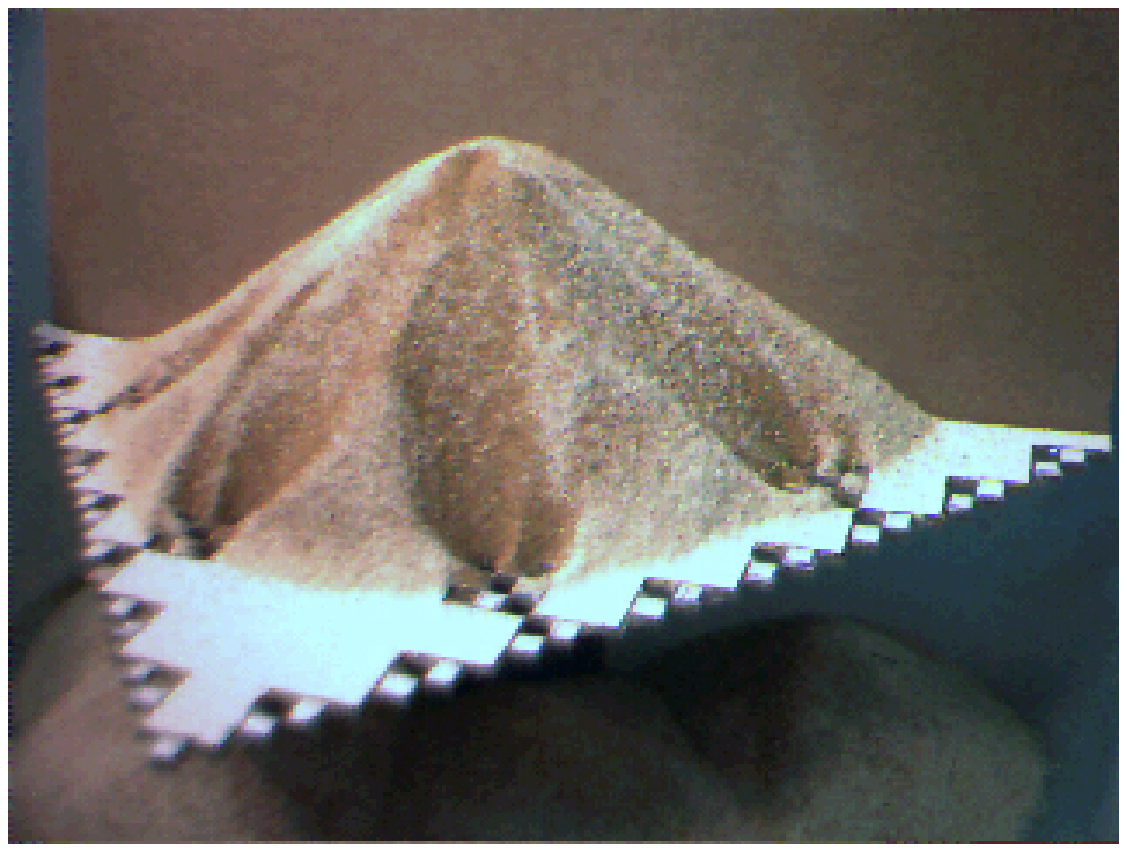,angle=90}
\end{picture}
\newpage

\begin{picture}(21,27)(-50,300)
\epsfig{file=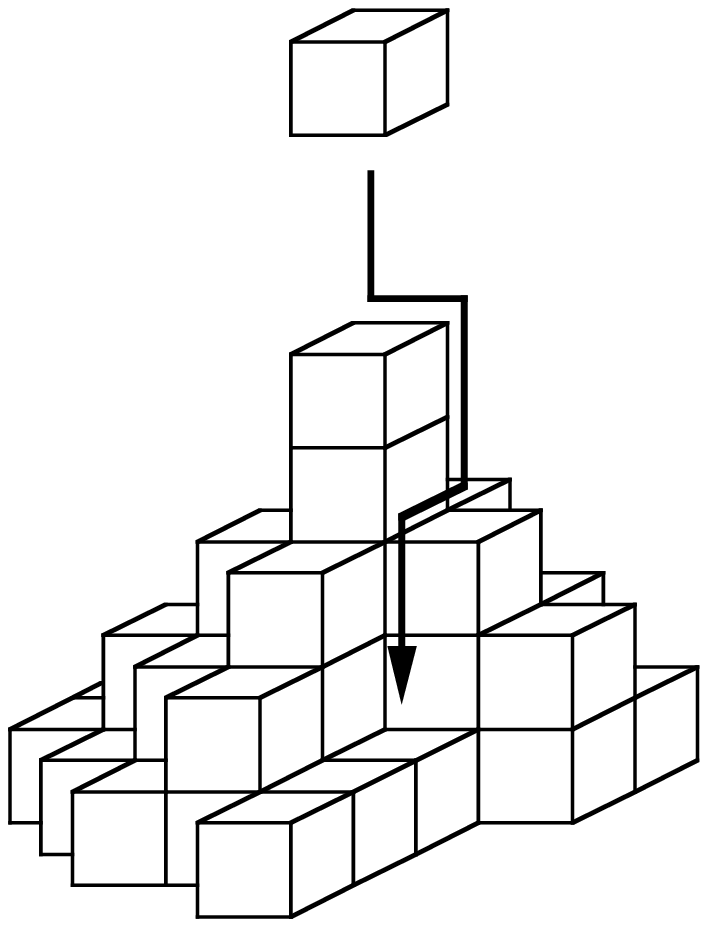}
\end{picture}
\newpage

\begin{picture}(21,27)(-50,300)
\epsfig{file=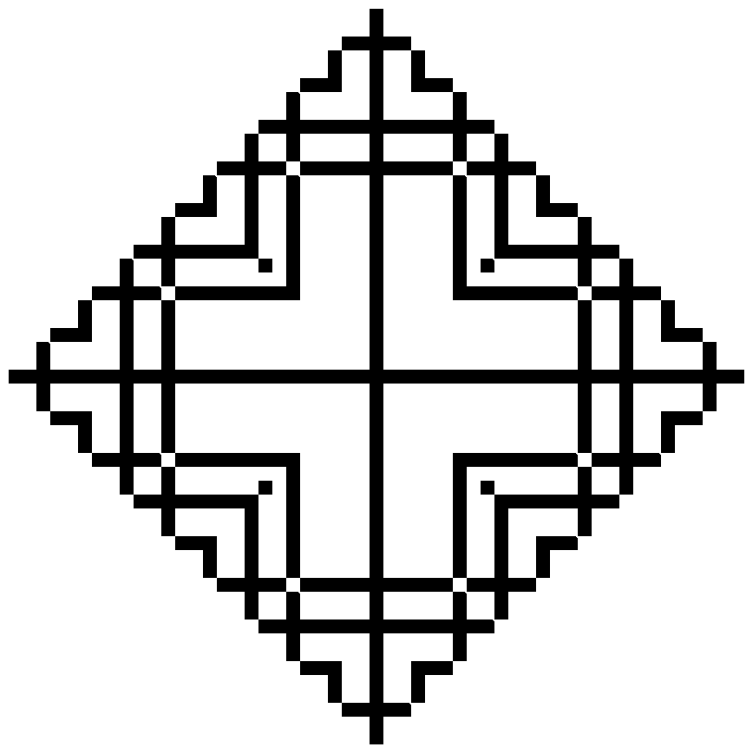}
\end{picture}
\newpage

\begin{picture}(21,27)(-50,300)
\epsfig{file=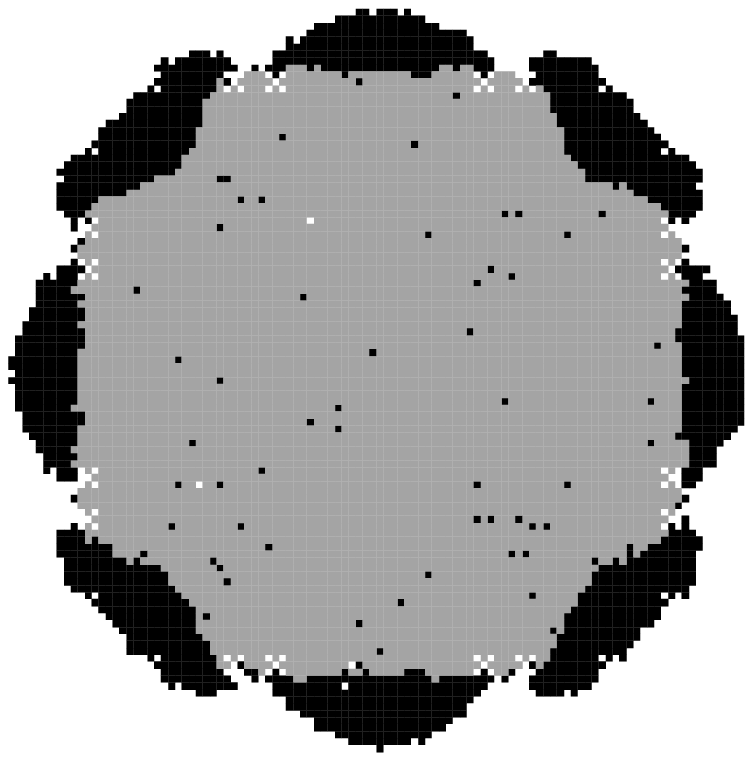}
\end{picture}
\newpage
\end{document}